\documentclass[aps, prl, amsmath, amssymb, verbatim]{revtex4}

\usepackage{graphicx}
\usepackage[dvips]{hyperref}
\usepackage{textcomp}
\usepackage[OT1,T1]{fontenc}

\begin{document}

\title{Is quantum mechanics based on an invariance principle?}
\author{L\'eon Brenig}
\affiliation{Universit\'e Libre de Bruxelles (ULB). CP.231. 1050 Brussels.
Belgium.}\label{I1}
\date{27 October 2006}

\begin{abstract}
Non-relativistic quantum mechanics for a free particle is shown to emerge from classical mechanics through an invariance principle under transformations that preserve the Heisenberg position-momentum inequality. These transformations are induced by isotropic space dilations. This invariance imposes a change in the laws of classical mechanics that exactly corresponds to the transition to quantum mechanics. The Schr\"odinger equation appears jointly with a second nonlinear equation describing non-unitary processes. Unitary and non-unitary evolutions are exclusive and appear sequentially in time. The non-unitary equation admits solutions that seem to correspond to the collapse of the wave function. \end{abstract}
\newcommand\revtex{Rev\TeX}
\maketitle{}

\subsection{I. Introduction}

Quantization of classical mechanics is generally not considered as deriving from an invariance principle. While relativity requires the invariance of the laws of nature under space-time transformations, quantum mechanics is usually presented as deriving from prescriptions relating classical quantities to Hermitian operators acting on Hilbert space.  The former theory is deeply rooted in space-time geometry, the latter is not. This deep difference is perhaps one of the main obstacles hampering the construction of a coherent theoretical framework for quantum gravity. \\
 
In contrast with this state of matter, very few attempts have been made to investigate the possibility that quantum mechanics could be derived from an invariance principle or, more generally, from space-time geometry. Among these works, the most relevant are the theory developed by L.Nottale  based on a fractal space-time and the principle of scale relativity, and the approach of E.Santamato and C.Castro relying on a Weyl geometry for space-time. \\

  L.Nottale and collaborators \cite{Nottale89},\cite{Nottale93},\cite{Nottale 04} are developing a general theoretical framework in which, as said above, space-time is supposed to have a fractal geometry. A second fundamental axiom of this theory is a generalization of the relativity principle to the scale transformations: The laws of Nature must be valid in every coordinate systems, whatever their state  of motion or of scale. We share completely this second axiom in our work, though, our implementation of it is different.\\
  Let us dwell more in details on the L.Nottale's theory. The assumption of a fractal structure of space-time reflects the fact that trajectories of elementary quantum particles are of fractal dimension two. This corresponds to the property first discovered by R.Feynman \cite{Feynman} that quantum trajectories, if one take their existence for granted, are of fractal dimension two. The non-differentiability of the trajectories on such a fractal space results in the existence of two velocity vectors at each point of the trajectory, the forward and backward tangent vectors. Their very existence permits a derivation of the Schr\"odinger equation using a scheme that is reminiscent of  Nelson's stochastic mechanics \cite{Nelson}. This is natural as the Brownian motion on which stochastic mechanics is based generates trajectories that are also of fractal dimension two. Such trajectories could indeed reflect the fractality of the space that bears them instead of resulting from a succession of random collisions. The non-differentiability of space and the bi-velocity structure of trajectories that follows from it lead to the introduction by L.Nottale of a scale-covariant complex time derivative. This operator is a complex combination of the forward and backward derivatives associated to Brownian diffusion. The replacement of usual time derivatives by scale-covariant ones in the laws of classical mechanics generates the quantum laws and the Schr\"odinger equation. This method is not limited to non-relativistic quantum mechanics but works also for the Klein-Gordon \cite{Nottale5} and the Dirac equations \cite{Nottale6}. It also provides interesting results in quantum field theory and high energy physics \cite{Nottale7}.\\
 As already said, scale relativity corresponds to the invariance of the laws of physics under scale transformations \cite{Nottale93} linking observers with different resolutions or scale states.   In the Nottale's theory, these transformations act on each couple of variables that are a physical field and its anomalous dimension. These variables are transformed under a dilatation or contraction of the observer's resolution. Improving a demonstration of the Lorentz transformations proposed by J.M.Levy-Leblond \cite{L-L}, L.Nottale assumes that it can also be  applied to relativity of scale and obtains an explicit form for the scale transformations. A strong consequence of these transformations is the prediction of an absolute and invariant minimum limit for lengths and times which is tentatively related to the Planck scale. Comparisons of these transformations and their consequences with those proposed in our work are discussed in the sequel of the present article.\\
 
  The other main geometric approach to  quantum mechanics is that developed mainly by E.Santamoto \cite{Santamato1}, \cite{Santamato2}, C.Castro \cite{Castro} and other researchers. Their theory is based on the assumption that space obeys a Weyl geometry. This, briefly said, corresponds to the hypothesis that the length of a vector whose origin is displaced parallely to itself along an arbitrary curve in such a space, changes along its path. Such spaces are not flat and are characterized by their Weyl curvature. The approach of E.Santamato and C.Castro involves from the start a probabilistic ingredient by assuming an initial statistical ensemble of positions for the particle. The dynamical law for a free particle is then derived  from a variational problem associated to a functional which is essentially the expectation over the position probability ensemble of the classical Lagrangian plus a supplementary term representing a coupling to the Weyl curvature of the space. As a result, the change in length of a parallely displaced vector can be related to the gradient of the logarithm of the position probability density.  The above authors are, then, able to show that the quantum potential \cite{Bohm} is proportional to the Weyl scalar curvature of the space. This, in turn, leads to an elegant derivation of the Schr\"odinger equation. \\
  
  The two theories described above involve a common element: The importance they ascribe to scale transformations for understanding quantum mechanics. This is also a main aspect of the work presented in this paper. \\ 

  Let us now turn to the results of the present article. The theory we propose here also relies on scale transformations between observers with different precisions and assumes the invariance of the laws of Nature under these transformations. Observers or frames of references are characterized not only by the origins of their space and time coordinates, the relative direction of their respective axis, their relative velocities but also by the relative accuracy or resolution of their measurements. Precision of measurements  is, consequently, embodied in geometry and the laws of nature must be invariant under precision scale transformations. In other words, quantum mechanics is viewed here as a kind of relativity theory under scale transformations, like in the Nottale's theory. Yet, in contrast with the latter, in our approach these transformations are simple homogeneous and isotropic dilatations of position variables , i.e. in the Nottale's terminology, they are "Galilean" scale transformations. They act as usual space dilatations on fields. Hence, in our approach a couple made of a field and its anomalous dimension does not undergo a Lorentzian-like scale transformation like those that are obtained in the work of L.Nottale.\\
    
  Our only requirement is the invariance of the Heisenberg inequality under position space dilatations.  As shown in the sequel, the action of these transformations on the classical definitions of the statistical uncertainties of the position and momentum of a free particle do not preserve the Heisenberg position-momentum inequality. As a consequence, we have to impose a modification of  the definition of  these uncertainties. Let us insist on the fact that this does not imply a change in the way the fundamental fields transform but, merely, in the way two global quantities, the statistical dispersions of position and momentum, that are functionals of these fields, transform. Since our description is based on fields - for a free particle these are its position probability density and its action-  defined on the position space, the only statistical moment that can be modified is the one characterizing the dispersion of momentum. Indeed, the only stochastic element assumed here concerns the position of the particle. Though, the quadratic position dispersion can be defined in an infinity of ways, all of these definitions must be homogeneous functionals of degree one of the position probability density only and must have physical dimension of the square of a length. These properties impose a unique transformation rule under spatial dilatations for the position dispersion as the transformation of the position probability density is constrained by the conservation of normalization. This is not the case for the momentum dispersion as momentum, in position space, is a derived quantity. Hence, the only statistical quantity that can be modified in order to keep the Heisenberg inequality invariant under dilations is the momentum uncertainty.
  This leads to a deep change in the dynamical law. Indeed, since the quadratic momentum dispersion is a linear function of the kinetic energy, its modification entails a change in the definition of the kinetic energy of the particle. As explained below and in the next chapter, the modification is uniquely determined and consists in a supplementary term which happens to be exactly the quantum potential \cite{Bohm}, thereby leading to a derivation of the Schr\"odinger equation.\\
  
  However, this is not the unique result. Our theory does not only recover the unitary dynamical evolution generated by the Schr\"odinger equation: It  also provides a non-unitary and nonlinear evolution equation for the wave function. This equation belongs to a large family of nonlinear Schr\"odinger equations known as the Doebner-Goldin family of equations \cite{Doebner}. The system of both Schr\"odinger and the new equations,  is invariant under scale transformations provided time is also transformed in a specific way. At first sight, the non-unitary evolution seems to unfold in a time variable that is different from that of the unitary evolution. However, it is argued that the two types of evolutions can only appear sequentially for a particle and, consequently, the two time parameters are the same but the natures of the two kinds of dynamical processes are fundamentally different. Though more work is needed, we present some reasons to believe that the non-unitary dynamics corresponding to the new equation could be related to processes like the collapse of the wave function.  Arguments in favor of this interpretation are the followings. First, the  non-unitary equation can be exactly linearized into a couple of forward and backward pure diffusion equations \cite{Auberson}. This corresponds rigorously to the so-called Euclidean quantum mechanics. Next, the system of forward and backward diffusion equations is shown to admit a class of solutions corresponding to a couple of prescribed initial and final conditions as indicated in an early article by E.Schr\"odinger \cite{Schrod} and more recently rigorously proved by J.C.Zambrini and collaborators \cite{Zambrini1}, \cite{Zambrini2}. Hence, among the solutions of this system there exists a subset of possible dynamical evolutions, the so-called Bernstein Markovian processes \cite{Bernstein}, starting from a specified initial state or wave function and reaching a reduced state corresponding to a measurement process. This is possible for the non-unitary evolution equation but not, of course, for the unitary, Schr\"odinger equation. In this scheme dynamical evolution could be seen as a succession of unitary and non-unitary processes respectively described by the two quantum equations.\\ 
     
  Before ending this Introduction, we should quote another important approach related to the question of the emergence of quantum mechanics, that of M.J.W. Hall and M. Reginatto \cite{Hall2}, \cite {Hall3}. This is even more necessary as we are using some important results of their work in our derivation. These authors assume that uncertainty is the essential property in which quantum and classical mechanics differ. This point of view leads them to postulate the existence of non-classical fluctuations of the momentum of a physical system. They assume, furthermore, that these fluctuations are entirely determined by the position probability density function. This enables them to derive the quantum dynamical law from the classical mechanics of a non-relativistic particle. To do so, they need two supplementary postulates that are causality and the additivity of the energy of N non-interacting particles. These last two postulates are also necessary in our derivation.\\ 
  Both their theory and ours allocate a fundamental importance to the Heisenberg uncertainty principle. Yet, the difference between the two approaches resides in the fact that the former needs to postulate the existence of non-classical momentum fluctuations and to assume that their statistical amplitude only depends on the position probability density. In contrast, in our work, these two postulates are derived from an invariance principle under scale transformations affecting the position and momentum uncertainties and preserving the Heisenberg inequality. These differences and similarities will be analyzed more deeply in the course of the present report. \\
  
  The course of this article is the following. In the second chapter, we introduce our main postulate  stating that the laws of Nature must be invariant under scale transformations. Among the laws of physics, the Heisenberg position-momentum inequality must be kept invariant by these transformations. We, then, deduce from that postulate the transformation rules of the position and momentum uncertainties.\\
   In the third chapter, we show that the classical mechanical definition of the momentum uncertainty is incompatible with these transformations. We are, thus, led to modify the classical definition of the momentum uncertainty in order to satisfy the imposed transformation rules. This modification is constrained by the transformations rules derived from our postulate and by the Hall-Reginatto conditions of causality and additivity of the kinetic energy of a system of non-interacting particles. This leads to a complete specification of the functional dependance of the supplementary term corresponding to the modification. The latter turns out to be proportional to the quantum potential. The passage from classical to quantum mechanics is, thus, fulfilled as the Schr\"odinger equation is a simple consequence of this result. \\
  The fourth chapter is devoted to the study of the variance under space dilations of the Schr\"odinger equation. It is shown that the latter is invariant jointly with another evolution equation for the wave function that is nonlinear and describes non-unitary processes in a new time parameter. Under a general space dilation and provided a specific transformation of the two times is performed, the Schr\"odinger equation and the new equation are invariant. \\
  In the fifth chapter we discuss the possible physical meaning of the nonlinear Schr\"odinger equation obtained in the precedent chapter. We first show that the time parameter associated to it is not independent from the usual time appearing in the linear Schr\"odinger: The evolutions respectively described by the linear and nonlinear equations must be successive. Hence, the only difference between the two times is a translation of their origins. Basing our arguments on an idea initiated by E.Schr\"odinger, we also show that the new equation admits a class of solutions that could represent processes of  wave function collapses. The chapter ends with general conclusions.

\subsection{II. Space dilatations and main postulate}

Let us consider a non-relativistic spinless free particle of mass $m$ in the 3-dimensional Euclidean space. In that space, observers are supposed to perform position measurements on the particle with instruments of limited precision. Hence, at a given instant the exact position of the particle is a random variable distributed with a probability density $\rho$(x). Limited precision on position measurement induces, in turn, uncertainty on momentum. Thus, an observer is characterized  by parameters denoting the statistical position and momentum uncertainties of its instruments. These parameters, let us call them  $\Delta x_k$ and $\Delta p_k$, for $k$ running from 1 to 3, are not uniquely defined as there exist many statistical measures of fluctuations for a given probability distribution. For example,  ${\Delta x_k}^{2}$  could be defined as the centered second moment of a given position probability density $\rho$(x) or as the  Fisher length \cite{Cox} associated to the same probability density. \\

In our picture, observers characterized by different values of their measurement uncertainties are related by space dilations. Our main postulate is the following: Under dilations of space coordinates, the laws of physics must be invariant. In particular, the Heisenberg position-momentum inequality 
\begin{equation}
{\Delta x_k}^{2}{\Delta p_k}^{2}\geqslant{\frac{\hbar^{2}}{4}}
\end{equation}

must be invariant for any of the three values of $k$.\\ 

This means that the parameters $\Delta x_k$ and $\Delta p_k$ must transform under spatial dilations in such a way that the above Heisenberg inequalities are kept invariant. In other words, the Heisenberg inequality is a fundamental invariant for the changes of precision relating all the observers, and precision becomes part of the geometrical description of the physical space.\\

To avoid proliferation of indices, we drop in the sequel the index k except in places where this would lead to an ambiguity. However, one must keep in mind that, when they appear in the same formulae or system of equations, $\Delta x$ and $\Delta p$ respectively represent  components labeled by the same value of the index $k$.\\

In order to implement the above postulate, let us now study its consequences. More precisely,
let us construct the transformation law that the quantity ${\Delta
x}^{2}{\Delta p}^{2}$ should undergo in order to fulfill the postulate.

Under an isotropic spatial dilatation of parameter $\alpha$,\ \ $x\rightarrow
e ^{-\alpha /2}x$, where $\alpha$ belongs to $ \mathbb{R}$, the
product ${\Delta x}^{2}{\Delta p}^{2}$ will transform as 
\begin{equation}
{\Delta x'}^{2}{\Delta p'}^{2}=f( {\Delta x}^{2}{\Delta p}^{2},
\alpha ) 
\end{equation}

In a succession of a dilatation of parameter $\alpha _{1}$ followed
by a second one of parameter $\alpha _{2}$, one should get
\begin{equation}
f( f( {\Delta x}^{2}{\Delta p}^{2}, \alpha _{1}) ,\alpha _{2}) =f(
{\Delta x}^{2}{\Delta p}^{2}, \alpha _{1}+\alpha _{2}) 
\end{equation}

Notice the additive character of the parameter $\alpha$. It results from
the additivity of this parameter in the spatial dilations.
This property also implies the commutativity of the two transformations of
parameters $\alpha _{1}$ and $\alpha _{2}$.

The identity transformation should continuously be reached when
taking the limit $\alpha$$ \rightarrow $0
\begin{equation}
{\lim }_{\alpha \rightarrow \text{$0$}}f( {\Delta x}^{2}{\Delta
p}^{2}, \alpha ) ={\Delta x}^{2}{\Delta p}^{2}
\end{equation}

 Furthermore, our postulate imposes the following condition
\begin{equation}
{\lim }_{\alpha \rightarrow \text{$\infty $}}f( {\Delta x}^{2}{\Delta
p}^{2}, \alpha ) =\frac{\hbar ^{2}}{4}
\end{equation}

for any values of ${\Delta x}^{2}{\Delta p}^{2}$$ \geqslant $$\frac{\hbar
^{2}}{4} $.

The above conditions amount to impose a one-parameter continuous
group structure for the set of these transformations along with
the existence of a fixed point, $ \frac{\hbar ^{2}}{4}$, for them.

These conditions are insufficient to characterize a unique form
for the function $f( {\Delta x}^{2}{\Delta p}^{2}, \alpha ) $. We
shall, henceforth, resort to introducing the following supplementary
but natural constraint. In the limit $ \hbar $$ \rightarrow $0,
the above transformation is expected to reduce to
\begin{equation}
{\Delta x'}^{2}{\Delta p'}^{2}= e ^{-2\alpha }{\Delta x}^{2}{\Delta
p}^{2}
\end{equation}

The reason for this form goes as follows. Under dilations $x\rightarrow
e ^{-\alpha /2}x$, the quantity ${\Delta x}^{2}$ , whatever the
choice made among its different possible definitions, as discussed
in the introduction, must transform as 
\begin{equation}
{\Delta x'}^{2}= e ^{-\alpha }{\Delta x}^{2}
\end{equation}

This law should not change in the limit $ \hbar $$ \rightarrow $0
as the definition of ${\Delta x}^{2}$ should not be affected by
the passage from classical to quantum mechanics as already discussed 
in the introduction. The above transformation of  ${\Delta x}^{2}$
comes from the fact that under a $x\rightarrow  e ^{-\alpha /2}x$
dilation, the position probability density $\rho$(x) transforms
as
\begin{equation}
\rho '\left( x\right) = e ^{\frac{3\alpha }{2}}\rho (  e ^{\frac{\alpha
}{2}}x) 
\end{equation}

This transformation law preserves the normalisation of the probability
density $\rho$ \cite{Hall}. Let us now consider the transformation law for ${\Delta
p}^{2}$. Remember that by this notation we denote the quadratic uncertainty of a given component of the vector $p$. The definition of this quantity for a classical particle
whose initial position is known only statistically via the position
probability density $\rho$ is given by
\begin{equation}
{{\Delta p}_{\mathrm{cl}}}^{2}=\int d^{3}x\\ \rho \\{\left( \partial
s\right) }^{2}-{\left( \int d^{3}x\\ \rho \\ \partial s
\right) }^{2}
\end{equation}

where $\partial$ denotes any component of the 3-D spatial gradient corresponding to the component of $\Delta p_{cl}$ we are considering in the above equation.
In equation (9),  $s$(x) represents the classical action of the particle. We shall consider
the following transformation of  $s$(x) under spatial dilatations
\begin{equation}
s'\left( x\right) = e ^{-\alpha }s(  e ^{\frac{\alpha }{2}}x) 
\end{equation}

This transformation of the action is justified by the fact that
the classical Hamilton-Jacobi equation for a free particle of mass
m and the continuity equation
\begin{gather}
\partial _{t}s =- \frac{{\left| \nabla s\right| }^{2}}{2m}
\\\partial _{t}\rho  = -\nabla .\left( \rho \frac{\nabla s}{m}\right)
\end{gather}

are kept invariant under isotropic dilations of space coordinates, $x'= e
^{-\alpha /2}x$, provided $s$(x) and $\rho$(x) transform as stated
above. The above transformations are different from those usually
considered in studies of the conformal invariance of the Hamilton-Jacobi
equation \cite{Barut}, \cite{Havas} where time is also dilated.

Turning back with these results to the transformation law for the
classical definition of ${{\Delta p}_{\mathrm{cl}}}^{2}$, we easily
find
\begin{equation}
{{\Delta p}_{\mathrm{cl}}'}^{2}= e ^{-\alpha }{{\Delta p}_{\mathrm{cl}}}^{2}
\end{equation}

This leads, of course to 
\begin{equation}
{\Delta x'}^{2}{{\Delta p}_{\mathrm{cl}}'}^{2}= e ^{-2\alpha }{\Delta
x}^{2}{{\Delta p}_{\mathrm{cl}}}^{2}
\end{equation}

The above reasoning justifies our previous constraint (6) on the limiting
form of the function $f( {\Delta x}^{2}{\Delta p}^{2}, \alpha )
$ when $ \hbar $$ \rightarrow $0. The dependence in ${\Delta x}^{2}{{\Delta
p}_{\mathrm{cl}}}^{2}$ in that limit is linear. Owing to this, we
shall assume a linear dependence on ${\Delta x}^{2}{\Delta p}^{2}$
for the function  $f( {\Delta x}^{2}{\Delta p}^{2}, \alpha ) $
\begin{equation}
{\Delta x'}^{2}{\Delta p'}^{2}=g( \alpha ) {\Delta x}^{2}{\Delta
p}^{2}+\frac{\hbar ^{2}}{4}k( \alpha ) 
\end{equation}

Though the linearity property introduced above is not completely
demonstrated by the previous reasoning, we use it for its simplicity
with the hope that further work will justify it completely. The
above conditions provide two coupled functional equations
for the two functions g($\alpha$) and k($\alpha$) 
\begin{gather}
g( \alpha _{1}) g( \alpha _{2}) =g(  \alpha _{1}+\alpha _{2}) 
\\g( \alpha _{2}) k( \alpha _{1}) +k( \alpha _{2}) =k( \alpha _{1}+\alpha
_{2}) 
\end{gather}

Using also the equation obtained from the permutation of indices
1 and 2 in the last equation along with the two equations above, we are led to the following result
\begin{equation}
{\Delta x'}^{2}{\Delta p'}^{2}= e ^{-\mathrm{n\alpha }}{\Delta x}^{2}{\Delta
p}^{2}+\frac{\hbar ^{2}}{4}\left( 1- e ^{-\mathrm{n\alpha }}\right)
\end{equation}

The constant $n$ is then found to be equal to 2 by using the limiting
form (6) of the transformation for $ \hbar $$ \rightarrow $0  and we
get
\begin{equation}
{\Delta x'}^{2}{\Delta p'}^{2}= e ^{-2\alpha }{\Delta x}^{2}{\Delta
p}^{2}+\frac{\hbar ^{2}}{4}\left( 1- e ^{-2\alpha }\right) 
\end{equation}

 Now, with the transformations on ${\Delta x}^{2}$ and on ${\Delta x}^{2}$${\Delta p}^{2}$ found above one can easily derive the transformation rule for  ${\Delta p}^{2}$. Finally, our postulate leads to the following transformations for ${\Delta x}^{2}$ and ${\Delta p}^{2}$
\begin{gather}
{\Delta x'_k}^{2}= e ^{-\alpha }{\Delta x_k}^{2}
\\{\Delta p'_k}^{2}= e ^{-\alpha }{\Delta p_k}^{2}+\frac{\hbar ^{2}}{4}\left(
e ^{\alpha }- e ^{-\alpha }\right) \frac{1}{{\Delta x_k}^{2}}
\end{gather}

where the parameter $\alpha$ is any real number. Exceptionally, we have restored index k running from 1 to 3  in the above formulae in order to stress again the fact that these laws are defined component-wise.\\

 The group property of these transformations is easily verified. The asymptotic behaviour under these transformations is also readily checked:
When $\alpha$$ \rightarrow $+$ \infty $ one has ${\Delta
x'}^{2}{\Delta p'}^{2}$$ \rightarrow $$\frac{\hbar ^{2}}{4}$. If
${\Delta x}^{2}{\Delta p}^{2}$ is already equal to $\frac{\hbar
^{2}}{4}$ then the product ${\Delta x'}^{2}{\Delta p'}^{2}$ keeps
the value $\frac{\hbar ^{2}}{4}$ for any value of $\alpha$. For
$\alpha$$ \rightarrow $-$ \infty $, ${\Delta x'}^{2}{\Delta p'}^{2}$$
\rightarrow $+$ \infty $ for any value of ${\Delta x}^{2}{\Delta
p}^{2}$$ \geq $$\frac{\hbar ^{2}}{4}$.\\

These above remarkable properties bear some similarities with the Lorentz
transformations. In analogy with the fact that the velocity of light constitutes
an upper limit for the velocities of material bodies, the parameter
$\frac{\hbar ^{2}}{4}$ represents a lower limit for the product
of uncertainties ${\Delta x}^{2}{\Delta p}^{2}$. This product plays a role similar to velocity in the Lorentz transformations. Latter in the article, the analogy will appear even more striking.\\

At this level, a comparison with the scale relativity theory developed by L.Nottale can be made. Though, the fundamental scope of his theory and ours are the same, its implementation presents differences. The scale transformations laws of L.Nottale's theory are given in the formulae (6.8.1a) and (6.8.1b) of reference \cite{Nottale93}. Formula (6.8.1a) concerns the dilatation ratio while (6.8.1b) concerns a scale-dependent field and its anomalous dimension in the sense of the renormalization group theory. The transformation (6.8.1a) gives the composition of two successive dilations, while (6.8.1b) is a scale transformation that can be applied to the position vector which, itself,  can be treated as a field with its own anomalous dimension. These transformations are not identical to the classical dilations $x\rightarrow e ^{-\alpha /2}x$ used in our approach. The latter correspond to "Galilean" scaling transformations in the Nottale's terminology. Moreover, the couple of variables that mix up in the "Lorentzian" scale transformations of L.Nottale are a field and its anomalous dimension. In contrast, the couple of variables that are mixed up in our scale transformations are the uncertainties associated to a couple of classically canonical variables that are position and momentum. \\

Let us come back to the account of our results. As we already mentioned, the definitions of ${\Delta x}$ and of ${\Delta p}$ as functionals of $s$(x) and $\rho$(x) are still unspecified. Their functional forms are derived in the next chapter from the condition that they transform under spatial dilations as postulated above.

\subsection{III. Recovering the quantum law of dynamics}

We now show that our postulate of the fundamental role of transformations (20), (21) 
imposes a radical modification of the laws of dynamics that precisely corresponds to the passage from classical to quantum mechanics.\\ 

To do so, let us start from the classical mechanical description of a free non-relativistic
particle of mass $m$ in the 3-D Euclidean space. As we did in the previous chapter, in order to take into account from the beginning the finite precision of the observer, we introduce an ensemble of initial positions characterized by the probability density $\rho$(x) . This function together with the classical action of the particle, $s$(x), are the basic variables of the formalism.  They are fields and due to this classical mechanics appears here as a field canonical theory \cite{Sudarshan83}. Let us stress that by assuming an initial position probability density we introduce only classical fluctuations of the position variable in the theory.\\

The time evolution of any functional of type
\begin{equation}
\mathcal{A} = \int d^{3}x F( x, \rho , \nabla \rho , \nabla \nabla
\varrho , ..., s, \nabla s, \nabla \nabla s, ...) 
\end{equation}

of the two variables $ \rho $ and $s$ and their spatial derivatives, at least once functionally
differentiable in terms of $ \rho $ and $s$, is given by
\begin{equation}
\partial _{t}\mathcal{A} = \left\{ \mathcal{A},\mathcal{H}_{\mathrm{cl}}\right\}
\end{equation}

where
\begin{equation}
\mathcal{H}_{\mathrm{cl}}= \int d^{3}x\ \ \frac{\rho {\left| \nabla
s\right|}^{2}}{2m}
\end{equation}

is the classical Hamiltonian functional and
\begin{equation}
\left\{ \mathcal{A},\mathcal{B}\right\}  = \int d^{3}x  \left[\frac{\delta
\mathcal{A}}{\delta \rho ( x) }\frac{\delta \mathcal{B}}{\delta
s( x) }- \frac{\delta \mathcal{B}}{\delta \rho ( x) }\frac{\delta
\mathcal{A}}{\delta s( x) }\right] 
\end{equation}

where $\frac{\delta }{\delta \rho( x) }$ and $\frac{\delta }{\delta
s( x) }$ are functional derivatives. The above functional Poisson
bracket endows the set of functionals of type (22) with an infinite
Lie algebra structure $ \mathbb{G}$.\\

Any functional belonging to $ \mathbb{G}$, and $\mathcal{H}_{\mathrm{cl}}$
is one of them, generates a one-parameter continuous group of transformations.
The time transformations are generated by $\mathcal{H}_{\mathrm{cl}}$.
Equation (23) when applied to $\rho$(x) and $s$(x) respectively, yields
the continuity equation and the Hamilton-Jacobi equation
\begin{gather}
\partial _{t}\rho  = -\nabla .\left( \rho \frac{\nabla s}{m}\right)
\\\partial _{t}s =- \frac{{\left| \nabla s\right| }^{2}}{2m}
\end{gather}

where the gradient $\nabla s$ is the classical momentum of the particle. It 
is a random variable over the ensemble of initial conditions corresponding to $\rho$(x).\\

Now let us consider the group of space dilatations $x$$\rightarrow$$e^{-\frac{\alpha
}{2}}$$x$ and its action on $\rho $ and $s$
\begin{equation}
\rho '\left( x\right) = e ^{\frac{3\alpha }{2}}\rho (  e ^{\frac{\alpha
}{2}}x) , \ \ s'\left( x\right) = e ^{-\alpha }s(  e ^{\frac{\alpha
}{2}}x) 
\end{equation}

where $\alpha$ is any real number. We have already discussed these transformations in the previous chapter. The important point to keep in mind is that they keep the dynamical equations (26) and (27) invariant.\\

To simplify the description, let us assume that the average momentum of the particle is vanishing.
This corresponds to a particular choice of a "comoving"  frame of reference
but, by no means, reduces the generality of our results. The general results
can, indeed, be retrieved by performing an arbitrary Galilean transformation.
In this particular frame, the classical definition of the quadratic uncertainty for a given component $k$ of the momentum is given by
\begin{equation}
{{\Delta p}_{\mathrm{cl,k}}}^{2}=\int d^{3}x\\ \rho \\( \partial_{k}
s)^{2}
\end{equation}

We now drop the index k , as in chapter II.
Under transformations (28), ${{\Delta p}_{\mathrm{cl}}}^{2}$ transforms as
\begin{equation}
{{\Delta p}_{\mathrm{cl}}'}^{2}= e ^{-\alpha }{{\Delta p}_{\mathrm{cl}}}^{2}
\end{equation}

Also, as discussed earlier, any definition of the  scalar quadratic position uncertainty measuring the dispersion of $\rho $(x), ${\Delta x}^{2}$, transforms component-wise as
\begin{equation}
{\Delta x'}^{2}= e ^{-\alpha }{\Delta x}^{2}
\end{equation}

Here, as in equations (20) and (21), the quantity ${\Delta x}^{2}$ still remains unspecified. 
Not surprisingly,  equation (30) shows that the classical quadratic
momentum uncertainty does not obey the transformation rule (21) prescribed by our postulate. As already discussed in chapter II, the transformation law (30) corresponds to the first term in the right hand side of equation (21). 
As a consequence, the requirement of transformations (20) and (21) to be fundamental compels us to modify the definition (29) of  ${{\Delta p}_{\mathrm{cl}}}^{2}$ in order to get a
quantity ${\Delta p}^{2}$ whose variance satisfies equation (21). This equation involves the 
constant ${\hbar ^{2}}$. Moreover, the new definition of ${\Delta p}^{2}$ should 
reduce to the classical one when ${\hbar}$ tends to zero. Indeed, in classical
mechanics the changes in the accuracy of position and momentum measurements
are not constrained by the Heisenberg inequality. This is clear when considering
equation (14). 
The desired modification to the definition (29) should, thus, consists in adding a supplementary 
term proportional to ${\hbar ^{2}}$. Indeed, the new quantity ${\Delta p}^{2}$ must
transform under isotropic dilations $x\rightarrow
e ^{-\alpha /2}x$ and laws (8), (10) in such a way that, at least, the first term of the right hand side of equation (19) is retrieved.

Let us translate this constraint by adding a new term to the above
definition of ${{\Delta p}_{\mathrm{cl}}}^{2}$ and get a new expression for the quadratic momentum uncertainty which, from now on, we shall denote by ${{\Delta p}_{q}}^{2}$
\begin{equation}
{{\Delta p}_{q,k}}^{2}=\int d^{3}x\\ \rho ( x) \left(\partial_{k} s( x)
)^{2} + \hbar ^{2}\mathcal{Q}_{k}\right. 
\end{equation}

where the index $k$ runs from 1 to 3. We shall drop this index k in the sequel and restore it only when it is necessary for clarity.\\

We now impose the condition that the spatial dilations and rules (8), (10)
should transform the quantity ${{\Delta p}_{q}}^{2}$
as prescribed by equations (20), (21), and prove that this reduces the set of possible
functional forms of $ \mathcal{Q}$. \\
First, note that following definition (29), the sum of the quadratic uncertainties (29) for the three components of the classical momentum is proportional to the classical energy functional (24). 
This is due to our choice of a comoving inertial reference frame. It is natural to consider that this
proportionality is preserved for the new definition of the quadratic momentum
uncertainty  ${{\Delta p}_{q}}^{2}$ we are looking for. It is also reasonable to assume
that the energy functional should belong to the Lie algebra $ \mathbb{G}$.
Hence, the new term $ \mathcal{Q}$ must also be a functional belonging
to the Lie algebra $ \mathbb{G}$, that is, it must be of the form
(22). This conclusion is, of course, valid for the three components $Q_k$.

Let us now apply the dilatation with the rules (8) and (10) to the definition (32)
of  ${{\Delta p}_{q}}^{2}$. This leads to
\begin{equation}
{{\Delta p'}_{q}}^{2}= e ^{-\alpha }{{\Delta p}_{\mathrm{cl}}}^{2}+
\hbar ^{2}\mathcal{Q}^{'}
\end{equation}

 where $\mathcal{Q}^{'}$ is the transform of $ \mathcal{Q}$ . \\
 Adding and subtracting an appropriate term, 
$e ^{-\alpha }\hbar ^{2}\mathcal{Q}$, to the right hand side of equation
(33) and using again definition (32), we get
\begin{equation}
{{\Delta p'}_{q}}^{2}= e ^{-\alpha }{{\Delta p}_{q}}^{2}+\hbar ^{2}(
\mathcal{Q}^{'}- e ^{-\alpha }\mathcal{Q}) 
\end{equation}

The identification of this equation with equation (21) imposes
\begin{equation}
\mathcal{Q}^{'}- e ^{-\alpha }\mathcal{Q}=\frac{1}{4{\Delta x}^{2}}\left(
e ^{\alpha }- e ^{-\alpha }\right) 
\end{equation}

which, using equation (20), can be transformed into
\begin{equation}
\mathcal{Q}^{'}-\frac{1}{4{\Delta x}^{'2}}= e ^{-\alpha }( \mathcal{Q}-\frac{1}{4{\Delta
x}^{ 2}}) 
\end{equation}

This equation possesses an infinity of solutions. However, its form
indicates the existence of a relation between $ \mathcal{Q}$ and
${\Delta x}^{ 2}$ that is scale independant
\begin{equation}
\mathcal{Q}_{k}=\frac{1}{4{\Delta x_k}^{ 2}}
\end{equation}

where the index k has temporally been restored.
This particular solution is the only one for which the relation
between ${{\Delta p}_{q}}^{2}$ and ${\Delta x}^{ 2}$ is independant
from the scale exponent $\alpha$.  Furthermore,  the sum of ${{\Delta p}_{q,k}}^{2}$
for the three values of the index $k$ should be proportional to the Hamiltonian functional, the generator of dynamics. One would, thus, expect that the latter keeps the same form
in term of ${\Delta x}^{ 2}$, independently of $\alpha$. In other
words, an observer should not be able to infer the value of $\alpha$
by doing only internal measurements of motion. This argument justifies
the choice of solution (37) on physical ground.

We are, thus, led to the conclusion that the supplementary
term necessary to obtain a definition of ${{\Delta p}_{q}}^{2}$ that
is compatible with the dilations and  (20), (21) is inversely proportional
to ${\Delta x}^{ 2}$, equation (37).  As this quantity only depends on
the probability density $\rho$(x), it is obvious that $ \mathcal{Q}$
must be a functional of the form (22) that does not depend on the
action $s$(x) or any of its spatial derivatives.

One should keep in mind, at this level, that the precise definition
of the quadratic position uncertainty, ${\Delta x}^{ 2}$, that appears
in transformations (20),  (21) and in relation (37) is still undetermined at this level. This
ambiguity is now lifted by considering the work of M.J.W. Hall and M. Reginatto \cite{Hall2}, \cite{Hall3} already mentioned in the Introduction. Their fundamental 
statement is the following. In order to explain the transition from classical
to quantum mechanics they assume that the classical momentum 
$\nabla s$(x) is affected by non-classical fluctuations represented by an additional random
variable of zero average and without correlation with $\nabla s$(x).
As a consequence, the scalar quadratic momentum uncertainty  contains
the classical term ${{\Delta p}_{\mathrm{cl}}}^{2}$ plus a correction
representing the quadratic average of the above fluctuations. Let
us stress that this is equivalent to our addition of a supplementary term  
$ \hbar ^{2}\mathcal{Q}$ in equation (32), although, the reason invoked by Hall
and Reginatto for adding this new contribution is completely different from ours. In
our approach this term comes from the necessity for the quadratic
momentum uncertainty to obey the transformation law (21) under dilatations, while in
the Hall-Reginatto theory, fluctuations are just postulated to exist. More precisely, the trace
of the statistical covariance of their fluctuations corresponds to the sum of our supplementary terms $Q_k$ for the three values of the index $k$ .
The next step in the Hall-Reginatto derivation is the assumption that this additive
term is only determined by the uncertainty in position, i.e. it only
depends functionally on $\rho $(x). Moreover, this term is assumed to behave
like the inverse of ${\Delta x}^{2}$ under dilatations. These two
last assumptions constitute what they call the exact uncertainty
principle. By comparison, in our approach these two assumptions are derived from the 
requirement for ${{\Delta p}_{q}}^{2}$ to transform as equations (20)
and (21) under space dilatations and from the requirement that the value of $\alpha$ could not be known by an observer by using only measurements made in his own frame of reference. 
  
At this level both our supplementary term $\mathcal{Q}$ and the quadratic average of the Hall-Reginatto's fluctuations have the same characteristics. We, thus, can now follow the rest of the Hall-Reginatto reasoning in order to get a complete determination of the functional expression of this term. To do so, they require two more principles that are very natural. Let us summarize them. The first one is causality. As we already stressed, the quadratic momentum uncertainty is related to the energy functional which, in turn, is the generator of dynamical motion. In our comoving frame of reference, this amounts to
\begin{equation}
\mathcal{H}_{q}=\operatorname*{\overset{3}{\sum }}\limits_{k=1}\frac{{{\Delta
p}_{q,k}}^{2}}{2m}
\end{equation}

The causality condition means that the equations of motions generated
by $\mathcal{H}_{q}$ should be causal, i.e. the existence and unicity
of their solutions should require only the specification of $\rho$(x)
and $s$(x) on an initial surface. This condition, combined with the exact uncertainty principle, enables Hall and Reginatto to show that $ \mathcal{Q}$ should only depend on the first order space derivatives of $\rho$(x).

The second principle required in the Hall-Reginatto theory is the so-called
independance condition, in other words, the Hamiltonian of N non-interacting
particles must be the sum of N terms. Each of these terms represents
the kinetic energy of a particle and only depends on the variables
of that particle.

Using these principles, Hall and Reginatto are able to prove  that the
unique functional form for $ \mathcal{Q}_{k}$ is
\begin{equation}
\mathcal{Q}_k=\beta \int d^{3}x{\left( \partial_k {\rho ( x) }^{1/2}\right)
}^{2}
\end{equation}

where $k$ runs from 1 to 3. Next, the constant $\beta$ is shown to be equal to one in order
to find, using equation (38), the quantum Hamiltonian functional which in the variables
$\rho$(x) and $s$(x) reads
\begin{equation}
\mathcal{H}_{q}=\int d^{3}x \left[\frac{\rho ( x) {\left| \nabla
s( x) \right|}^{2} }{2m} +\frac{ \hbar ^{2}}{2m}{\left| \nabla {\rho
( x) }^{1/2}\right| }^{2}\right] 
\end{equation}

Simultaneously, we obtain the complete determination of ${\Delta x_k}$ that appears in equations (20) and (21) by using the relations (37) and (39). Interestingly, what is obtained is not the usual definition corresponding to the second order centered statistical moment of the component k of the position vector $x$. The definition obtained here is, up to a numerical factor, proportional to the classical Fisher length \cite{Hall}, \cite{Cox} associated to the position probability density $\rho$(x).

The functional $\mathcal{H}_{q}$ generates the quantum time evolution
of any functional $\mathcal{A} $ of the algebra $ \mathbb{G}$ via
equation (23) where $\mathcal{H}_{\mathrm{cl}}$ is to be replaced
by $\mathcal{H}_{q}$. When $\mathcal{A}$ is specialized to $s$(x)
an easy calculation leads to a modified Hamilton-Jacobi equation
\begin{equation}
\partial _{t}s =- \frac{{\left| \nabla s\right| }^{2}}{2m}+ \frac{
\hbar ^{2}}{2m}\frac{\nabla ^{2}\rho ^{1/2}}{\rho ^{1/2}}
\end{equation}

while the continuity equation for $\rho$(x), equation (26), is preserved.
The supplementary term appearing in the Hamilton-Jacobi equation can be recognized as the so-called quantum potential \cite{Bohm}. Due to the presence of this typically quantum contribution, the Schr\"odinger equation is readily obtained from equation
(41) and the continuity equation (26) by performing the transformation
from the variables $\rho$(x) and $s$(x) to the wave function variables
$\psi$ and $\psi ^{*}$
\begin{equation}
\psi =\rho ^{1/2} e ^{\mathrm{is}/\hbar }
\end{equation}

Notice that in the algebra defined by the Poisson bracket (25), the above transformation is canonical.\\

Let us summarize. We have derived the quantum evolution law for a free
non-relativistic spinless particle in 3-D flat space from the requirement that the quadratic 
uncertainties on position and momentum should satisfy the transformations rules (20) and (21)
together with the causality and independance principles. The form
in which we obtain quantum mechanics is that of the canonical field
theory which has been introduced and studied from different points
of view by various authors \cite{Strocchi}, \cite{Heslot}, \cite{Guerra83},
\cite{Ashtekar}. None of these authors, however, derives quantum
mechanics from an invariance principle as we do here.

\subsection{IV. Scale invariance and the non-unitary evolution equation} 

Let us now consider the variance of the Schr\"odinger equation under
the spatial dilatations and transformation laws (8), (10). By adding and substracting adequate terms, the transformation of the Hamiltonian functional (40) under these
transformations can be cast in the explicit form
\begin{multline}
{\mathcal{H}'}_{q}[ \rho ,s,\nabla \rho , \nabla s] =\mathrm{cosh\alpha
}\int d^{3}x\left[\frac{\rho ( x) {\left| \nabla s( x) \right|}^{2}
}{2m} +\frac{ \hbar ^{2}}{2m}{\left| \nabla {\rho ( x) }^{1/2}\right|
}^{2}\right] -\\
\mathrm{sinh\alpha }\int d^{3}x\left[\frac{\rho ( x) {\left|
\nabla s( x) \right|}^{2} }{2m} -\frac{ \hbar ^{2}}{2m}{\left| \nabla
{\rho ( x) }^{1/2}\right| }^{2}\right] 
\end{multline}

where  ${\mathcal{H}'}_{q}$, as a functional of $\rho$(x),
$s$(x) and their respective spatial derivatives, is obtained from
\begin{equation}
{\mathcal{H}'}_{q}[ \rho ,s,\nabla \rho , \nabla s] \equiv  \mathcal{H}_{q}[
\rho ' ,s',\nabla \rho  ',\nabla s'] 
\end{equation}

in which $\rho ' ,s',\nabla \rho  ',\nabla s'$ are derived from
equations (8), (10).

The first term in the right hand side of equation (43) is proportional
to $\mathcal{H}_{q}[ \rho ,s,\nabla \rho , \nabla s] $ while the
second term contains a factor that is similar to $\mathcal{H}_{q}[
\rho ,s,\nabla \rho , \nabla s] $ up to a sign in the integral. Let us call $\mathcal{K}_{q}$
this factor
\begin{equation}
\mathcal{K}_{q}[ \rho ,s,\nabla \rho , \nabla s] \equiv \int d^{3}x\left[\frac{\rho ( x) {\left| \nabla s( x) \right|}^{2} }{2m} -\frac{
\hbar ^{2}}{2m}{\left| \nabla {\rho ( x) }^{1/2}\right| }^{2}\right]
\end{equation}

The physical dimension of $\mathcal{K}_{q}$ is clearly the same as that of 
$\mathcal{H}_{q}$, i.e. it is an energy. As any functional belonging to the 
algebra $ \mathbb{G}$, $\mathcal{K}_{q}$ is the generator of a one parameter continuous
group. Let us denote by $\tau$ the parameter of that group. Since
$\mathcal{K}_{q}$ has the dimension of an energy, the dimension
of $\tau$ is that of a time. In terms of this new functional, the
transformation (43) can be rewritten in a more compact notation as
\begin{equation}
{\mathcal{H}'}_{q}=\mathrm{cosh\alpha }\ \ \mathcal{H}_{q}-\mathrm{sinh\alpha
}\ \ \mathcal{K}_{q}
\end{equation}

while $\mathcal{K}_{q}$ can easily be shown to transform as
\begin{equation}
{\mathcal{K}'}_{q}=-\mathrm{sinh\alpha }\ \ \mathcal{H}_{q}+\mathrm{cosh\alpha
}\ \ \mathcal{K}_{q}
\end{equation}

Notice that these transformations strictly derive from equations (20), (21).

Hence, under the dilatations and transformations (8), (10), the couple ($ \mathcal{H}_{q}$, $\mathcal{K}_{q}$)
transforms as a 2-D Minkowski vector under a Lorentz-like transformation.
One easily shows that this induces the following transformations
on the group parameters $t$ and $\tau$ respectively associated to $\mathcal{H}_{q}$
and\ \ $\mathcal{K}_{q}$
\begin{equation}
t'=\mathrm{cosh\alpha }\ \ t+\mathrm{sinh\alpha }\ \ \tau 
\end{equation}

and
\begin{equation}
\tau '=\mathrm{sinh\alpha }\ \ t+\mathrm{cosh\alpha }\ \ \tau 
\end{equation}

Now, any functional $ \mathcal{A}$ of the algebra $ \mathbb{G}$ can be considered
as a function of both $t$ and $\tau$, and its evolution in both times variables
is given by
\begin{equation}
\partial _{t}\mathcal{A} = \left\{ \mathcal{A},\mathcal{H}_{q}\right\}
\end{equation}

and
\begin{equation}
 \partial _{\tau }\mathcal{A} = \left\{ \mathcal{A},\mathcal{K}_{q}\right\}
\end{equation}

Let us perform a dilation transformation of parameter $\alpha$ 
on equations (50) and (51). A simple calculation yields 
\begin{equation}
\partial _{t''}\mathcal{A}' = \left\{ \mathcal{A}',{\mathcal{H}'}_{q}\right\}
\end{equation}

and 
\begin{equation}
\partial _{\tau ''}\mathcal{A}' = \left\{ \mathcal{A}',{\mathcal{K}'}_{q}\right\}
\end{equation}

where  $t''$ and $\tau''$ correspond to  the rescaling of  $t'$ and $\tau'$ by a factor  $e^{-\alpha}$
\begin{equation}
t''=e ^{-\alpha }{\left(\mathrm{cosh\alpha }\ \ t+\mathrm{sinh\alpha }\ \ \tau \right)}
\end{equation}
and
\begin{equation}
\tau ''=e ^{-\alpha }{\left(\mathrm{sinh\alpha }\ \ t+\mathrm{cosh\alpha }\ \ \tau \right)}
\end{equation}

The necessity of rescaling the time variables comes from the fact that the spatial dilation and laws (8), (10) do not constitute a canonical transformation in the sense of the Poisson bracket (25). This is related to the non-conservation of the action in this transformation. The canonical character is restored by the above time rescaling.
In other words, we have proven that the equations of evolutions generated by both Hamiltonian functionals are covariant under transformations (8), (10) provided their respective time parameters are transformed as prescribed by equations (54), (55). 

The Schr\"odinger equation is a particular case of equation (50)
for 
\begin{equation}
\mathcal{A}=\psi =\rho ^{1/2} e ^{\mathrm{is}/\hbar }
\end{equation}

and the calculation of the Poisson bracket leads to the usual form
\begin{equation}
\mathrm{i\hbar }\partial _{t}\psi =-\frac{\hbar ^{2}}{2m}\nabla
^{2}\psi 
\end{equation}

Now, the wave function, $\psi$, can also be considered as a function
of $\tau$. Its evolution equation in this parameter is easily derived
from equation (51) and reads
\begin{equation}
\mathrm{i\hbar }\partial _{\tau }\psi =-\frac{\hbar ^{2}}{2m}\nabla
^{2}\psi +\frac{\hbar ^{2}}{m}\psi \frac{\nabla ^{2}\left| \psi
\right| }{\left| \psi \right| }
\end{equation}

We shall discuss the possible physical interpretation of this equation in the next chapter. \\

As a result of the above results, the system of equations (57) and
(58) is covariant under the space dilatations and its transform reads
\begin{gather}
\mathrm{i\hbar }\partial _{t''}\psi '=-\frac{\hbar ^{2}}{2m}\nabla
^{2}\psi '
\\\mathrm{i\hbar }\partial _{\tau ''}\psi '=-\frac{\hbar ^{2}}{2m}\nabla
^{2}\psi '+\frac{\hbar ^{2}}{m}\psi '\frac{\nabla ^{2}\left| \psi
'\right| }{\left| \psi '\right| }
\end{gather}

where the transformation of the wave function
\begin{equation}
\psi '\left( x\right) ={{{ e ^{\frac{3\alpha }{4}}[ \psi (  e ^{\frac{\alpha
}{2}}x) ] }^{\frac{1+ e ^{-\alpha }}{2}}}{[ \psi ^{*}(  e ^{\frac{\alpha
}{2}}x) ] }^{\frac{1- e ^{-\alpha }}{2}}}
\end{equation}

directly derives from the dilatation laws (8), (10). The nonlinearity of transformation (61)
is remarkable and contrasts with the linear transformation
rules that generally are assumed in the studies of invariance groups of the
Schr\"odinger equation \cite{Havas}, \cite{Barut}, \cite{Niederer}. The reason for that difference clearly appears when considering among others the article by P. Havas \cite{Havas}. In this work, the transformation rules of both the classical Hamilton-Jacobi and the Schr\"odinger equations under spatial dilatations and, more generally under the conformal group, are studied. When considering the transformation of the Hamilton-Jacobi equation, the classical action $s$ is supposed to transform as prescribed in our equation (10). However, when the transformation of the Schr\"odinger equation under dilatations is considered, only a restricted form of this transformation is considered leading to the fact that the $\psi$ function transforms as the square root of a density, i.e. as ${\rho}^{\frac{1}{2}}$. This hypothesis does not take into account the fact that $\psi$, as given by equation (56), is a function of both ${\rho}^{\frac{1}{2}}$ and $s$. As $s$ in the quantum case obeys a modified Hamilton-Jacobi equation (41), there is no reason to assume that this quantity does not transform under dilatations. The reason to discard the transformation of  $s$ in the wave function in the above mentioned studies is unclear but it is perhaps related to the fact that this quantity appears in $\psi$ via a complex phase factor of modulus one. However, there is no fundamental argument that can support this hypothesis when the Schr\"odinger equation is decomposed in terms of the continuity equation (26) and the modified Hamilton-Jacobi equation (41).

Before ending this article, another approach to the transformations (46), (47) should be mentioned. This was in fact the first we considered chronologically. These transformation rules
can, indeed, be generated by the following element of the algebra $ \mathbb{G}$
whose definition is
\begin{equation}
\mathcal{S}=\int d^{3}x \ \ \rho ( x) s( x) 
\end{equation}

It represents the average on the position ensemble of the classical action or, up to a factor
$\hbar $, the ensemble average of the quantum phase.

An easy calculation using the definition (25) of the functional Poisson
bracket gives
\begin{equation}
\left\{ \mathcal{S},\mathcal{H}_{q} \right\} =\int d^{3}x \left[\frac{\rho
{\left| \nabla s\right|}^{2}}{2m} -\frac{ \hbar ^{2}}{2m}{\left|
\nabla \rho ^{1/2}\right| }^{2}\right] =\mathcal{K}_{q}
\end{equation}

and 
\begin{equation}
\left\{ \mathcal{S},\mathcal{K}_{q} \right\} =\int d^{3}x \left[\frac{\rho
{\left| \nabla s\right|}^{2}}{2m} +\frac{ \hbar ^{2}}{2m}{\left|
\nabla \rho ^{1/2}\right| }^{2}\right] =\mathcal{H}_{q}
\end{equation}

The infinitesimal transformation for the parameter $\delta \alpha$ generated by 
$ \mathcal{S}$ of any element $ \mathcal{A}$ of the algebra $ \mathbb{G}$ is defined as
\begin{equation}
\mathcal{A}'=\mathcal{A}+\delta \alpha \left\{ \mathcal{A},\mathcal{S}\right\}
\end{equation}

Let us apply (65) respectively to both $\mathcal{H}_{q}$ and $\mathcal{K}_{q}$.
It is easily shown that after exponentiating these infinitesimal transformations in order to generate
the transformation for finite values of $\alpha$ one
recovers equations (46) and (47). As a consequence, transformations (20) and (21) are also recovered.

Note also that both generators $\mathcal{H}_{q}$ and $\mathcal{K}_{q}$ tend
to $\mathcal{H}_{\mathrm{cl}}$ for $\hbar $$ \rightarrow $0, i.e.
the times evolution in $t$ and $\tau$ become identical in the classical
limit. Moreover, the transformation of the time variables (54), (55) become the identity transformation for the unique time parameter.
This seems to indicate that the finiteness of $\hbar $ is lifting a degeneracy
that is intrinsical to classical mechanics, and splits the two time variables or as we shall argue in the next chapter, splits two family of processes of different natures.

Another remarkable property that can be derived from the above
relations is the fact that $\mathcal{H}_{q}+i$$\mathcal{K}_{q}$
is a holomorphic function of $t$+i$\tau$.

\subsection{V. Discussion of the nonlinear Schr\"odinger equation and conclusions}

The nonlinear Schr\"odinger equation (58) in the variable $\tau$,
obtained here as a companion to the usual linear Schr\"odinger equation
in the time $t$, is not a newcomer in physics. It has been postulated,
though in the time $t$ variable and in different contexts, by several
authors \cite{Guerra 1}, \cite{Vigier}, \cite{Smolin}. It belongs
to the class of Weinberg's nonlinear Schr\"odinger equations \cite{Weinberg}.
This equation admits a nonlinear superposition principle \cite{Auberson}.
It has been studied as a member of the general class of nonlinear Schr\"odinger equations generated by the so-called nonlinear gauge transformations introduced by Doebner
and Goldin \cite{Doebner}. The evolution generated by this equation
in the $\tau$ variable is nonunitary as $\mathcal{K}_{q}$ can not
be reduced to the quantum average of a Hermitian operator. In addition,
one easily shows that together with the functionals of the algebra $ \mathbb{G}$ generating translations, rotations and Galilean boosts, $\mathcal{K}_{q}$ constitutes a functional
canonical representation in $ \mathbb{G}$ of the Galilei algebra. This means that equation (58) is Galilean invariant. Another important property is that equation (58) also implies the continuity equation for the probability density function $\rho$. Hence, though non-unitary,
this equation obeys minimal  physical requirements such as Galilean invariance and
the equation  of continuity .\\

What is the physical meaning of equation (58) and of the temporal
parameter $\tau$? In relation with this question, it is intriguing
to notice that, for a free particle, in the $\tau$ evolution the product ($\partial
_{\tau }$${\Delta x}^{2})$($\partial _{\tau }{\Delta p}^{2}$) is
always negative. This is reminiscent of the process of state vector
reduction in position measurement in which ${\Delta x}^{2}\rightarrow
0$ while ${\Delta p}^{2}\rightarrow +\infty $, or conversely if
one is measuring momentum. Would this $\tau$ evolution be related
in some way to the nonunitary process that physicists like R.Penrose
\cite{Penrose} are trying to identify for the description of the
wave function collapse? We present now some arguments going in that direction.\\

First, let us discuss the physical meaning of the second time variable,
$\tau$. The calculation of the crossed time derivative of any functional
$ \mathcal{A}$\ \ gives
\begin{equation}
\left( \partial _{t}\partial _{\tau }-\partial _{\tau }\partial
_{t}\right) \mathcal{A}=\left\{ \left\{ \mathcal{H}_{q},\mathcal{K}_{q}\right\}
, \mathcal{A}\right\} 
\end{equation}

The right-hand side of the above equation is generally different
from zero for most functionals $ \mathcal{A}$. This means that the
two times can not be considered as two independent variables. An
example is given by the case where $ \mathcal{A}$  is ${\Delta
x}^{2}$, where a sum over k running from 1 to 3 is taken into account
in  ${\Delta x}^{2}$
\begin{equation}
\left( \partial _{t}\partial _{\tau }-\partial _{\tau }\partial
_{t}\right) {\Delta x}^{2}=\frac{8\hbar ^{2}}{m^{2}}\int d^{3}x
{\left| \nabla \rho ^{1/2}\right| }^{2}
\end{equation}

The only conclusion that can be drawn from this constatation is
that both time variables t and $\tau$ represent the same physical
time, however, the processes they parametrize are of different natures
and can not occur simultaneously for a given physical system. An
analogous situation would be the situation in which a particle is
submitted during a first lapse of time ${\Delta t}_{1}$ to an external
potential $V_{1}$ and, then, during a consecutive time interval ${\Delta
t}_{2}$ it is submitted to a different external potential $V_{2}$.

Clearly, the two time intervals could not overlap finitely. In the
opposite case, since the Hamiltonians corresponding to both potentials
are different and do not generally commute, we would be confronted
to the non commutativity of the crossed times derivative. In other
terms, a given system can not be submitted to different evolutions
simultaneously! This could seem obvious, but in our case, this sheds
another light on our results.

A non-unitary evolution processe generated by $\mathcal{K}_{q}$
is, thus, expected to follow or precede a unitary process governed
by $\mathcal{H}_{q}$.

Next, let us discuss the nature of these non-unitary processes.
They are solutions of the nonlinear Schr\"odinger equation (58).
It is known that this equation, and its complex conjugate, can be
exactly linearized \cite{Doebner}, \cite{Auberson}. Indeed,
in terms of the following two functions $\varphi$ and $\hat{\varphi
}$
\begin{gather}
\varphi =\rho ^{1/2} e ^{s/\hbar }
\\\hat{\varphi }=\rho ^{1/2} e ^{-s/\hbar }
\end{gather}

the system of equation (58) and its complex conjugate transform
in 
\begin{gather}
\hbar \partial _{\tau }\varphi =\frac{\hbar ^{2}}{2m}\nabla ^{2}\varphi
\\\hbar \partial _{\tau }\hat{\varphi }=-\frac{\hbar ^{2}}{2m}\nabla
^{2}\hat{\varphi }
\end{gather}

i.e., a forward and a backward diffusion equations. 

These equations are often considered as deriving from the usual
linear Schr\"odinger equation by replacing $t$ by $-it$. This
leads to what is called Euclidean Quantum Mechanics. \\

An interesting property of this system of equations is that, in
contrast with the usual  Schr\"odinger equation, it admits a class
of solutions corresponding to an initial function and a final function
that are prescribed. These are the so-called Bernstein diffusion
processes \cite{Bernstein}. This type of solutions have been first
contemplated by E.Schr\"odinger himself \cite{Schrod} for the
diffusion equation. He was, in fact, trying to see whether the Schr\"odinger
equation also could admit such solutions in order to explain the
paradoxes of the quantum coherence and of the wave function collapse.
However, the unitarity of the processes described by the Schr\"odinger
equation excludes such solutions. J.C.Zambrini and collaborators
\cite{Zambrini1}, \cite{Zambrini2} have clarified the status
of these solutions for the forward and backward diffusion equations.
They proved the existence and unicity of these solutions for any
couple of given well-behaved initial and final functions. \\

This leads us to conclude that processes like the wave function
collapse due to a measurement could belong to that class of Bernstein
solutions of the nonlinear equation (58). Indeed, in such a process
the initial state is specified, but the reduced final state is in
some sense prepared by the operation of measurement. This hypothesis 
is at the focus of our present investigation and its results will be exposed in
a forthcoming publication.\\

In conclusion, the requirement of covariance under space dilatations
that preserve the Heisenberg inequality leads not only to the unitary
processes described by the usual Quantum Mechanics, it generates
also an equation describing non-unitary processes that could correspond
to the collapse processes. Both types of processes can occur only
in succession and are coupled in the scale transformations corresponding
to our postulate. More work on this question is necessary and study
of experimental situations where this interpretation could be verified
will be carried out.\\

Another interesting question emerging from the above framework concerns
the consequences of requiring local invariance under the dilatations
(8), (10), i.e dilatations with space dependent parameter $\alpha$(x).
Would this requirement result in the existence of a new fundamental
interaction field?\\

Finally, the most exciting question is about the picture of space-time
that would emerge from the combination of the special or general
relativity invariance with the quantum invariance described here.\\

The author wants to dedicate this article to his master in Physics and friend
Dr.R.Balescu who suddenly deceased during this work and who encouraged
him in this approach. He wants also to thank Drs. C.George, J.Reignier,
M.J.W.Hall, Y.Elskens, I.Veretennicoff, G.Barnich, R.Lambiotte,
A.Saa, M.Vincke, V.Faraoni, M.Czachor and Mr.F.Ngo for their useful comments on this work.
Email address: lbrenig@ulb.ac.be

\end{document}